# Modulating Electronic Structure of Monolayer Transition Metal Dichalcogenides by Substitutional Nb-Doping

Lei Tang,[†] Runzhang Xu,[†] Junyang Tan,[†] Yuting Luo,[†] Jingyun Zou,[†] Zongteng Zhang,[‡] Rongjie Zhang,[†] Yue Zhao,[‡] Junhao Lin,[‡] Xiaolong Zou,[†] Bilu Liu[*†] and Hui-Ming Cheng[*†#]

[†]Shenzhen Geim Graphene Center, Tsinghua-Berkeley Shenzhen Institute & Tsinghua Shenzhen International Graduate School, Tsinghua University, Shenzhen 518055, P. R. China.

[‡]Department of Physics, Southern University of Science and Technology, Shenzhen 518055, P. R. China.

[#]Shenyang National Laboratory for Materials Science, Institute of Metal Research, Chinese Academy of Sciences, Shenyang 110016, P. R. China.

Correspondence should be addressed to bilu.liu@sz.tsinghua.edu.cn

hmcheng@sz.tsinghua.edu.cn

**ABSTRACT**: Modulating electronic structure of monolayer transition metal dichalcogenides (TMDCs) is important for many applications and doping is an effective way towards this goal, yet is challenging to control. Here we report the *in-situ* substitutional doping of niobium (Nb) into TMDCs with tunable concentrations during chemical vapour deposition. Taking monolayer $WS_2$ as an example, doping Nb into its lattice leads to bandgap changes in the range 1.98 eV to 1.65 eV. Noteworthy, electrical transport measurements and density functional theory calculations show that the 4$d$ electron orbitals of the Nb dopants contribute to the density of states of Nb-doped $WS_2$ around the Fermi level, resulting in an $n$ to $p$-type conversion. Nb-doping also reduces the energy barrier of hydrogen absorption in $WS_2$, leading to an improved electrocatalytic hydrogen evolution performance. These results highlight the effectiveness of controlled doping in modulating the electronic structure of TMDCs and their use in electronic related applications.

**INTRODUCTION**

Two-dimensional (2D) transition metal dichalcogenides (TMDCs) have been considered promising candidates for high-performance electronics, optoelectronics, and electrocatalysis,[1-6] however the controlled modulation of the electronic structure is a key challenge. It is known from silicon that the carrier type and structure can be changed by doping with boron ($p$-type) or phosphorus ($n$-type) with well-controlled concentrations.[7] Therefore, the doping of heteroatoms into TMDCs may effectively modulate their electronic structure such as carrier type,[8] which determines their physicochemical properties.[9-12] Meanwhile, for electrocatalysts, changing the local atomic configurations of electrocatalysts by doping may also improve reactions such as the hydrogen evolution reaction (HER)[13] and the oxygen evolution reaction.[14]

Great efforts have recently been made to tune the electronic structure of TMDCs, including charge transfer induced by the absorption of other molecules or functional groups,[15-17] plasma treatment,[18-20] and cation doping during the chemical vapor transport (CVT) growth of TMDCs.[21-22] Note that these methods have some deficiencies. For example, charge transfer doping is usually unstable with time, plasma treatment may damage the original structure of the TMDCs, and CVT is time-consuming and only bulk materials can be grown. Chemical vapor deposition (CVD) has been considered a more suitable approach for incorporating heteroatoms into 2D TMDCs during the growth process to overcome these problems, facilitating the steady, damage-free and efficient preparation of monolayer TMDCs with a controlled doping concentration. Indeed, the doping of different metal or chalcogen atoms into TMDCs during CVD growth has been reported in several systems.[23-29] For instance, Qin *et al.* have grown Nb-doped $WS_2$ by using liquid phase precursors in a two-step CVD method, while the as-grown samples have nonuniform distributions of PL intensity.[30] Li *et al.* reported doping Se into $MoS_2$, where Se and S are from the same group and provide limited modulation of the intrinsic properties of the TMDCs.[11] It is clear that more effort is needed in exploring the effective doping and related applications of TMDCs.

Here, we use a one-step CVD method for the doping of Nb atoms into the lattice of a monolayer TMDC to achieve controllable modulation of its electronic structure. We show that the bandgaps of monolayer $WS_2$ can be changed from 1.98 eV to 1.65 eV by the addition of a controlled concentration of Nb dopant from 0.3 at% to 4.7 at%. The Nb-doped $WS_2$ has a high crystallinity, a tunable composition and property, as well as good uniformity. Electrical transport measurements show that Nb-doping converts *n*-type $WS_2$ into *p*-type which has a lower overpotential for HER than pure $WS_2$, *i.e.*, 90 mV at 10 mA cm$^{-2}$ and 320 mV at 1000 mA cm$^{-2}$. Such a change in the electronic properties and device characteristics is explained by density functional theory (DFT) calculations, showing that the 4*d* electron orbitals of Nb dopant atoms contribute to the density of states (DOS) around the Fermi level in Nb-doped $WS_2$. These results together show the importance of substitutional doping in changing the properties of 2D TMDCs and expanding their applications that involve the electronic structure.

**RESULTS AND DISCUSSION**

We first studied the bandgap engineering of TMDCs by changing the dopant concentration, using Nb-doped monolayer $WS_2$ as an example. It is known that both $WS_2$ and $NbS_2$ have the 2H-phase crystal structure with similar lattice constants ($WS_2$ [a, b, c] = [3.16, 3.16, 12.47] Å, $NbS_2$ [a, b, c] = [3.32, 3.32, 11.94] Å),[31] and the covalent radii of W (137 pm) and Nb (134 pm) are very close. These features suggest the possible substitutional doping of Nb into $WS_2$ with few defects and the top and side view structures of Nb-doped $WS_2$ are shown in Figure 1a. The DFT results show that when Nb doping concentrations in $WS_2$ reach 2.0 at% and 4.0 at%, corresponding to one Nb substitutional dopant atom in a 2D 7-by-7 and 5-by-5 superlattice, respectively, they induce hole doping and increase the energy level of the valence band maximum (VBM) of $WS_2$ (Figure 1b). The positions of the Fermi level ($E_f$, taking the VBM energy level) change with the Nb doping concentration according to the DFT calculations (Figure 1c). Therefore, the substitutional doping of Nb atoms in monolayer $WS_2$ changes its electronic structure.

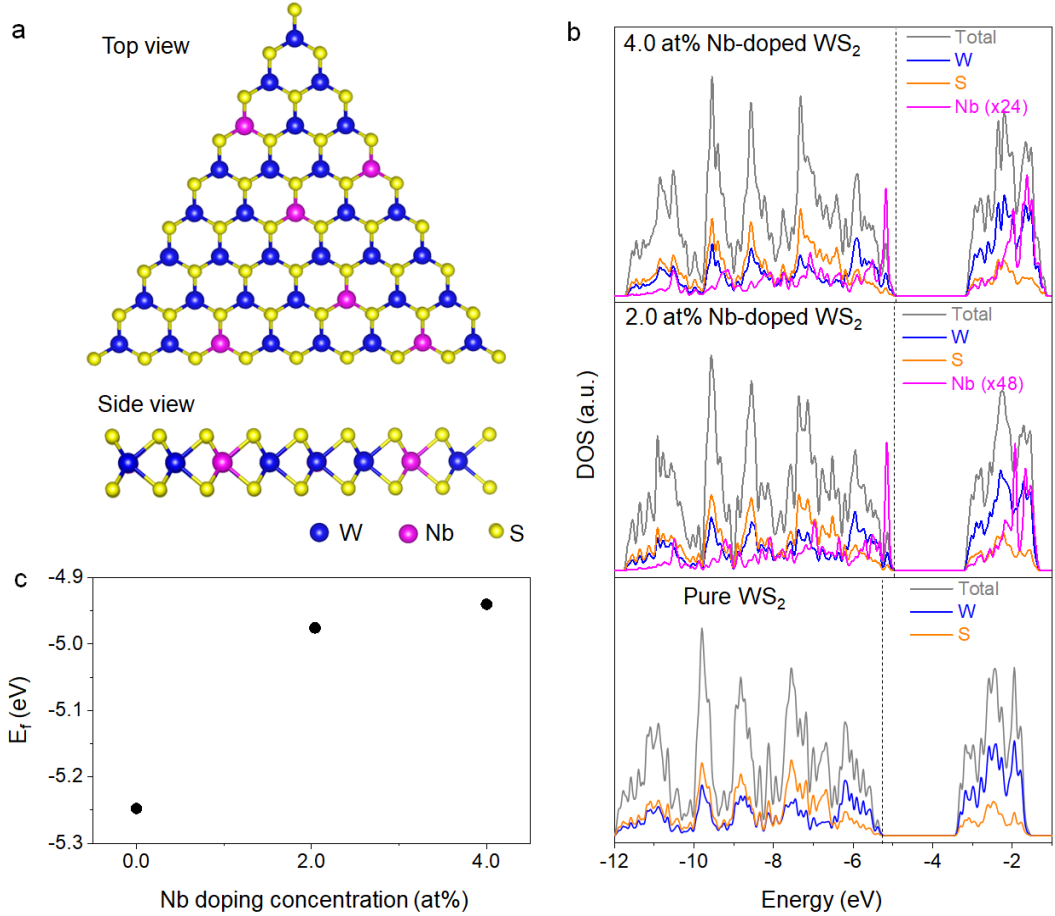

**Figure 1.** (a) Top and side view structures of Nb-doped $WS_2$. (b) DFT calculations of the DOS in pure $WS_2$ (bottom panel), 2.0 at% Nb-doped $WS_2$ (middle panel), and 4.0 at% Nb-doped $WS_2$ (top panel). The energy of the VBM is taken as the Fermi level denoted by grey dashed lines. (c) The relationship between the position of the Fermi level for different Nb doping concentrations.

Motivated by the DFT results, we prepared monolayer Nb-doped $WS_2$ with different doping concentrations using an *in-situ* substitutional doping method based on CVD. Briefly, mixtures of $WO_3$, $NbCl_5$, and NaCl as well as sulphur were used to grow Nb-doped $WS_2$ at 780 °C (Figure 2a, see details in the Experimental Section). The Nb-doped $WS_2$ has a triangular shape with a grain size of ~30 μm and a thickness of 0.62 nm (Figures 2b and 2c), showing the formation of monolayer samples. Pure $WS_2$ has a strong photoluminescence (PL) peak at 1.98 eV while he Nb-doped $WS_2$ with the highest

Nb concentration of 4.7 at% has a weak peak at 1.65 eV, which is red shifted by 330 meV compared with the pure WS$_2$ (Figures 2d and 2e). The Raman spectra (Figure 2f) of the Nb-doped and pure WS$_2$ show the same peaks at E$_{2g}^1$ and A$_{1g}$. Similar phenomena in terms of quenched PL and unchanged Raman spectra are shown in Nb-doped MoS$_2$ systems (Figures S1a and S1b). We obtained PL and Raman intensity maps (E$_{2g}^1$ and A$_{1g}$ peaks) of Nb-doped WS$_2$ (Figures 2g-i) as well as Nb-doped MoS$_2$ (Figures S1c-f) and they showed good homogeneity. X-ray photoelectron spectroscopy (XPS) shows peaks associated with the Nb 3$d$ core levels at 209.5 eV and 206.8 eV in the Nb-doped WS$_2$ (Figure 2j) and the core-level peaks of W and S show a shift to lower binding energies compared with the pure WS$_2$ (Figures 2k and 2l), indicating that the Fermi level of the Nb-doped WS$_2$ is shifted toward the VBM.[32] Overall, these results show that the Nb-doped WS$_2$ is homogeneous and its electron structure can be modified in a controlled manner by changing the amount of Nb dopant.

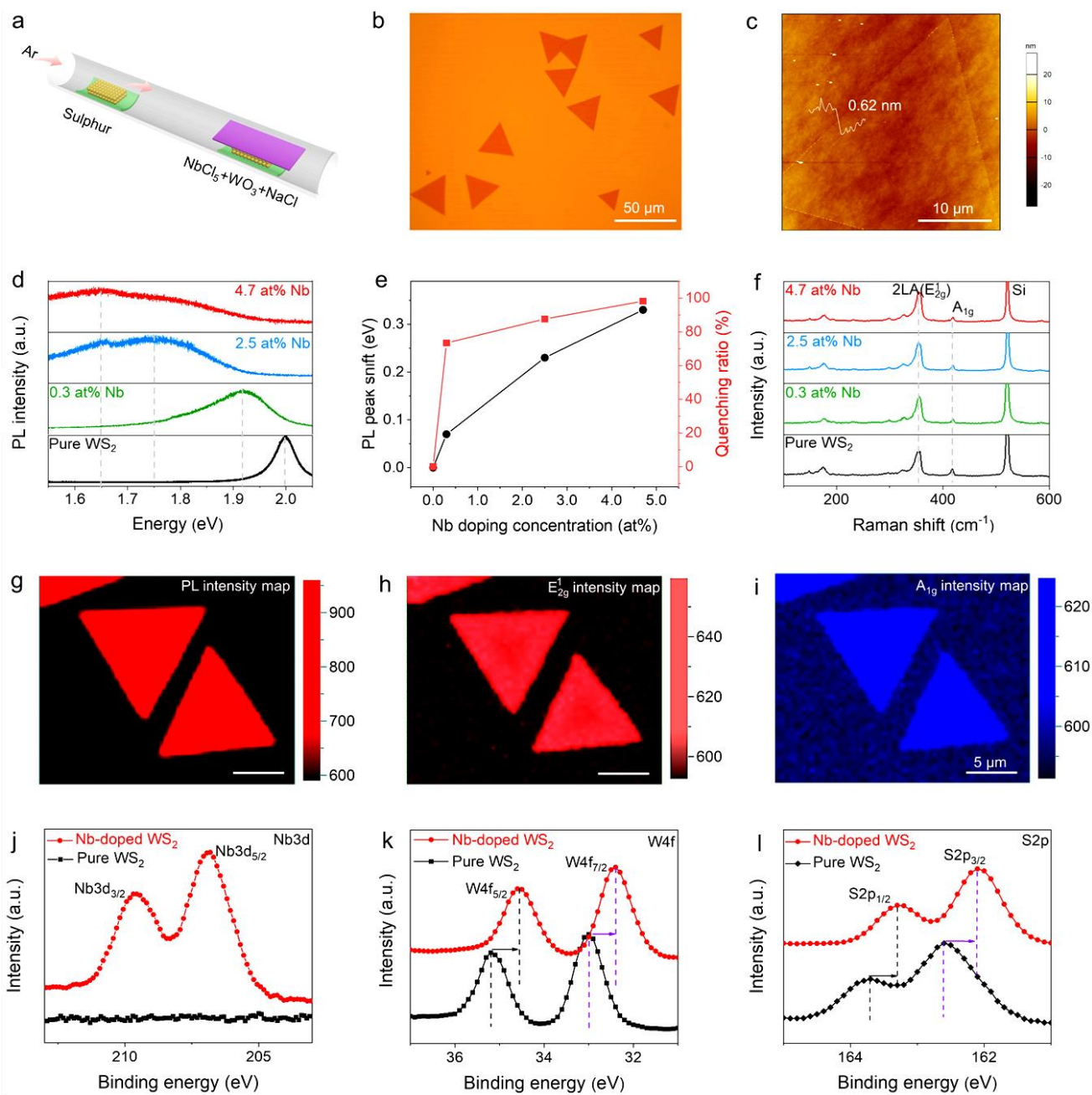

Figure 2. Growth and characterization of the Nb-doped WS$_2$. (a) Schematic of the CVD growth of Nb-doped TMDCs. (b, c) Typical optical and atomic force images of Nb-doped monolayer WS$_2$. (d-f) Comparisons of the PL and Raman spectra of the Nb-doped and pure WS$_2$. (g-i) PL (1.92 eV) and Raman (E$_{2g}^1$, 352 cm$^{-1}$; A$_{1g}$, 418 cm$^{-1}$) intensity maps of the Nb-doped WS$_2$. (j-l) Comparison of the XPS Nb, W, and S core levels of the Nb-doped and pure WS$_2$.

We then characterized the microstructures of the monolayer Nb-doped $WS_2$ by high-angle annular dark-field scanning transmission electron microscopy (HAADF-STEM). Compared with the pure $WS_2$ (Figure 3a), the darker contrast at the cation sites indicates substitutional Nb atoms due to its much smaller atomic number than W (Figures 3b-d). W vacancies can also be differentiated from the Nb dopant due to the almost absent contrast. The intensity ratio of Nb/W is ~0.4, as indicated by the experimental intensity profiles and agreed well with the simulation results (Figure 3e). The doping concentrations of Nb are calculated using atom-by-atom quantitative intensity mapping (Figures 3f-h). By increasing the amount of Nb precursor, the doping concentration of Nb increased from 0.3 at% to 4.7 at%. Energy-dispersive X-ray spectroscopy (EDX) shows that the Nb dopant atoms are uniformly incorporated in the $WS_2$ lattice (Figure S2). These results show that substitutional doping of Nb in $WS_2$ with different concentrations has been achieved.

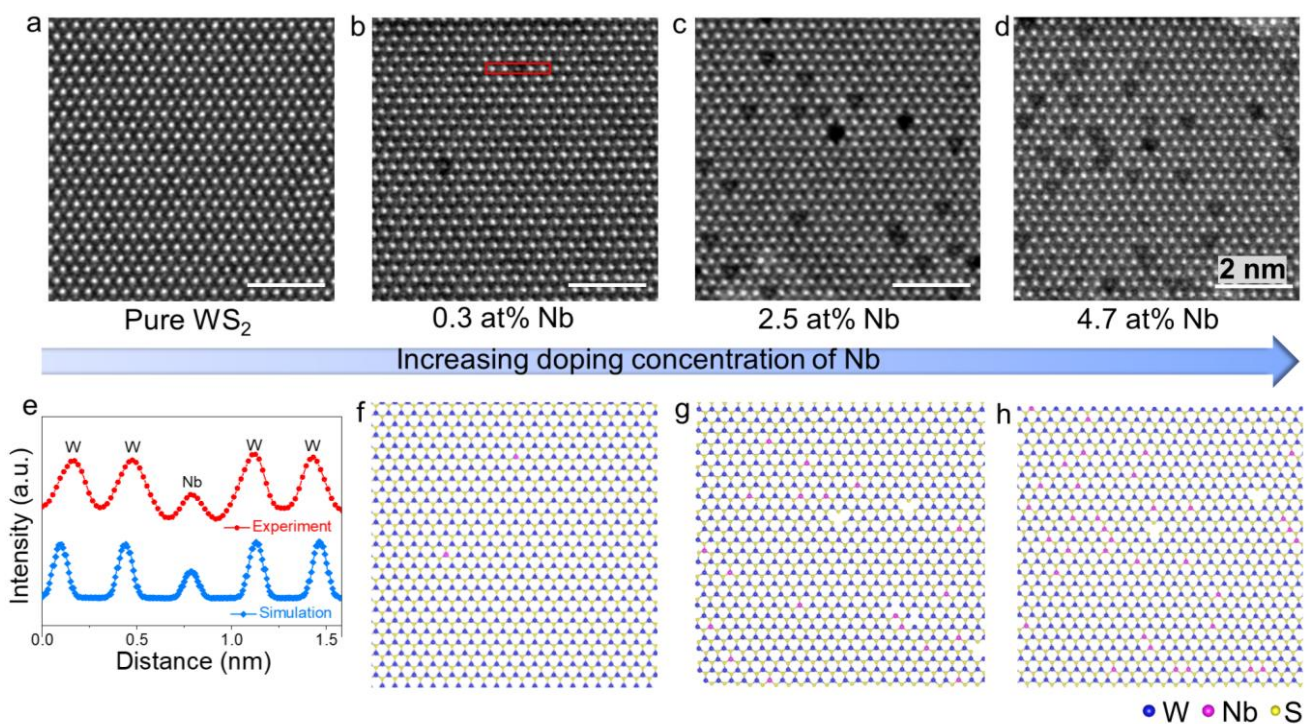

Figure 3. Microstructural characterization of the Nb-doped $WS_2$. (a-d) HAADF-STEM images of the pure and Nb-doped $WS_2$ with different doping concentrations. (e) Comparison of the experimental and simulated intensity profiles taken from the red box in Figure 3b, showing the typical W-Nb-W configuration. (f-h) Corresponding atom-by-atom maps show the positions of W (blue), Nb (pink), and S (yellow) atoms in Nb-doped $WS_2$.

To investigate the effect of Nb doping on changing the electronic structure of $WS_2$, we fabricated and measured the properties of field effect transistors (FETs) based on Nb-doped $WS_2$. The devices were fabricated by a direct laser writing technique followed by the deposition of Cr/Au as the drain and source contacts (details in the Experimental Section). A pure $WS_2$-based FET shows *n*-type conduction (black curve, Figure 4a), which is consistent with previous studies.[20, 24] Note that the FET based on Nb-doped $WS_2$ with a low doping concentration (2.5 at% Nb) shows ambipolar conduction behavior (blue curve, Figure 4a). For a high doping concentration (4.7 at% Nb), it shows *p* type conduction behavior (red curve, Figure 4a). The carrier concentrations (*n*) and threshold voltages ($V_{th}$) of these devices were also investigated based on the transfer curves (Figure S3) and the electron concentration ($n_e$) changed from $4.92 \times 10^{11}$ cm$^{-2}$ to $6.16 \times 10^{11}$ cm$^{-2}$ to $8.00 \times 10^{10}$ cm$^{-2}$ at $V_g = 80$ V, the hole concentration ($n_h$) changed from $1.03 \times 10^{12}$ cm$^{-2}$ to $6.62 \times 10^{12}$ cm$^{-2}$ at $V_g = -80$ V (Figure 4b), and the threshold voltage ($V_{th}$) changed from 67 V, 65 V to 14 V (Figure 4c) (see Experimental Section for the details of how these values were obtained). DFT calculations were used to understand these changes in electrical properties, and showed that the valence electronic states of pure $WS_2$ primarily originate from hybridization of the 5*d* orbitals from W and the 3*p* orbitals from S (bottom panel in Figure 1b). When Nb atoms are doped into the $WS_2$ lattice, they induce additional sharp defect states above the VBM of pure $WS_2$ (middle and top panels in Figure 1b), which serve as a shallow acceptor level. Overall, the results show that modulating the electronic structure of $WS_2$ contributes to the *n* to

*p*-type conversion.

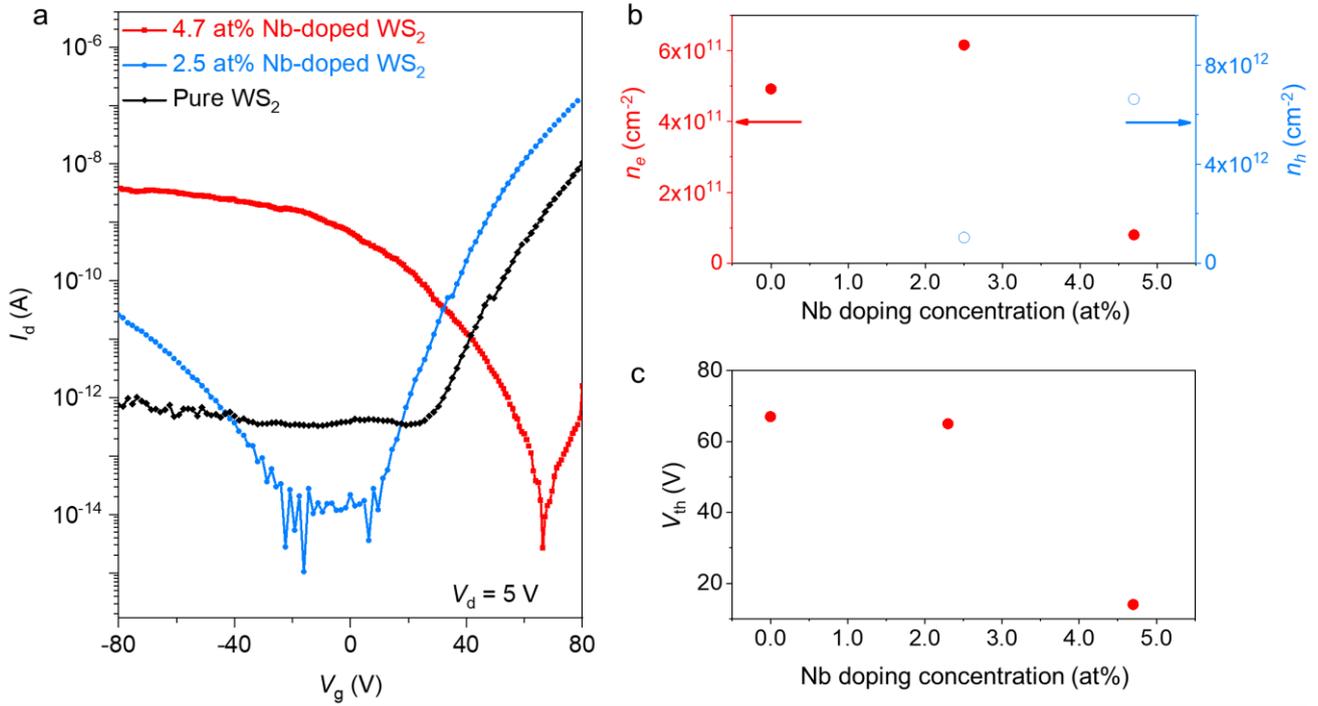

Figure 4. Transport properties of the Nb-doped WS$_2$. (a) Transfer characteristics of FETs fabricated using pure WS$_2$, 2.5 at% Nb-doped WS$_2$, and 4.7 at% Nb-doped WS$_2$. (b) Relationship between carrier concentration and Nb doping concentration. (c) Relationship between $V_{th}$ and Nb doping concentration.

To further explore the effect of Nb doping on changing the electronic structure of WS$_2$, we also checked the HER activity of Nb-doped WS$_2$ using a microreactor design.[33] After two rounds of electron beam lithography Cr/Au was deposited to serve as the drain and source contacts (Figures 5a and S4; details in the Experimental Section). Linear sweep voltammograms (LSV) were obtained on the pure and 4.7 at% Nb-doped WS$_2$ (Figure 5b) and the corresponding results of the overpotential at different current densities are summarized in Figure 5c, showing overpotentials of 90 mV at 10 mA cm$^{-2}$ and 320 mV at 1000 mA cm$^{-2}$. Obviously, Nb-doped WS$_2$ is more active for HER than pure WS$_2$. The Gibbs free energy of hydrogen adsorption ($\Delta G_H$) was calculated as an indicator for HER performance.[34] Taking all the active sites into consideration (Figure S5), our DFT results show that hydrogen adsorption at the in-plane center of the hexagonal ring next to the Nb dopant (the most preferable site, denoted as "hollow-in") has the lowest energy barrier (~0.11 eV) (Figures 5d-f). In addition, the 4.7 at% Nb-doped $WS_2$ reduces the energy barrier for valence states interacting with $H^+$, thus enhancing $H_2$ adsorption. Overall, these results show that modulating the electronic structure of $WS_2$ plays an important role in fast HER kinetics.

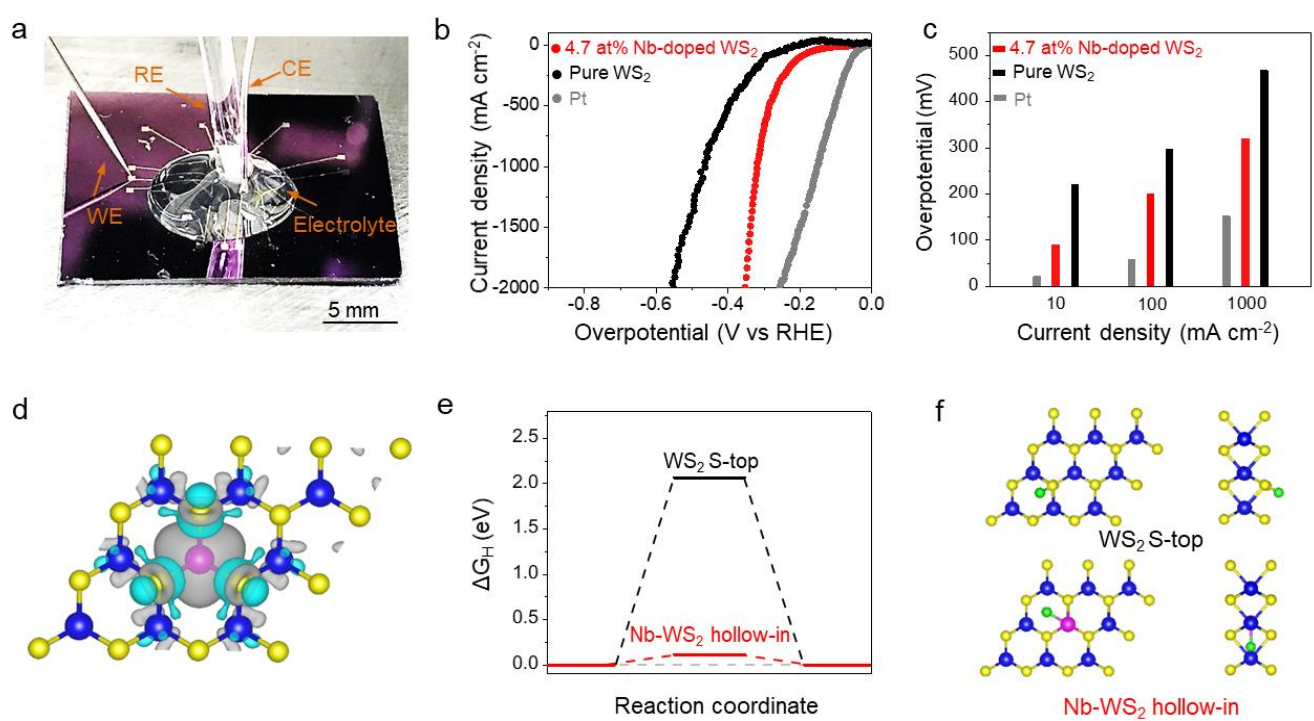

Figure 5. Improved HER performance of the Nb-doped $WS_2$. (a) The typical picture of the HER-microreactor device. (b) LSV curves of the microreactor devices based on pure $WS_2$ and 4.7 at% Nb-doped $WS_2$. (c) Summary of the overpotential at different current densities. (d) The charge density variance of $WS_2$ after substitutional Nb-doping. (e) $\Delta G_H$ versus the reaction coordinates for the most preferable $H^+$ adsorption sites in the pure and Nb-doped $WS_2$. (f) Corresponding schematic structures of these two most preferable sites. The blue, yellow, pink, and green spheres respectively denote W, S, Nb atoms, and $H^+$ ions.

**CONCLUSION**

We have modulated the electronic structure of monolayer TMDCs by the substitutional doping of Nb during the CVD process. Systematic characterization of the doped materials shows tunable compositions from 0.3 to 4.7 at% and a red shift of the optical bandgap from 1.98 eV to 1.65 eV. Moreover, the Nb-doped $WS_2$ shows an *n*- to *p*-type conversion behavior and a low overpotential for HER electrocatalysis (90 mV at 10 mA cm$^{-2}$ and 320 mV at 1000 mA cm$^{-2}$). Theoretical studies indicate that the 4*d* electron orbitals of Nb dopant atoms contribute to the DOS around the Fermi level in Nb-doped $WS_2$. Nb-doping also reduces the energy barrier of hydrogen absorption, leading to an improved HER performance. The study sheds light on the mechanism of bandgap engineering, which opens up further applications based on 2D TMDCs.

**EXPERIMENTAL SECTION**

*Materials and chemicals.* Sulphur powder (99.95%, Alfa Aesar, USA), niobium (V) chloride ($NbCl_5$, 99.9%, Shanghai Macklin Biochemical Co., Ltd., China), tungsten (VI) oxide powder ($WO_3$, 99.99%, Alfa Aesar, USA), molybdenum (VI) oxide powder ($MoO_3$, 99.99%, Alfa Aesar, USA), sodium chloride (NaCl, 99.5%, Shanghai Macklin Biochemical Co., Ltd., China), potassium hydroxide (KOH, 90%, Shanghai Macklin Biochemical Co., Ltd., China), sulphuric acid ($H_2SO_4$, 98wt%, Shanghai Macklin Biochemical Co., Ltd., China), polymethyl methacrylate (PMMA, Allresist, AR-P 679.04, Germany), $SiO_2$/Si substrate (300 nm oxide layer thickness, Hefei Kejing Materials Technology Co., Ltd., China), acetone, ethanol and isopropyl alcohol (Analytical Reagent, Shanghai Macklin Biochemical Co., Ltd., China) were used as-received. *CVD growth of Nb-doped $WS_2$, Nb-* *doped MoS$_2$, pure WS$_2$, and pure MoS$_2$*. The growth was conducted in a homemade two-zone CVD setup with a 1-inch diameter quartz tube (TF55035C-1, Lindberg/Blue M, Thermo Fisher Scientific, USA). Sulfur powder (100 mg) was placed in the first zone at upstream, WO$_3$ mixed with NaCl (mass ratio 6:1), and NbCl$_5$ with different masses (1 mg, 2 mg, 10 mg) with a facedown SiO$_2$/Si substrate were placed downstream (8-10 cm) for the growth of the Nb-doped WS$_2$. The tube was flushed with Ar for 30-40 min to remove the air and the furnace was then heated to 780-800 °C with a heating rate of 40 °C min$^{-1}$ and kept at this temperature for 1-5 min for the growth of Nb-doped WS$_2$ before being cooled to room temperature. A flow of 100 standard cubic centimeter per minute (sccm) of Ar was introduced during the whole process. The Nb-doped MoS$_2$, pure WS$_2$, and MoS$_2$ were grown using the same method, using MoO$_3$ mixed with NaCl (mass ratio 6:1), and NbCl$_5$ with different masses (2 mg, 10 mg), WO$_3$ and NaCl (mass ratio 6:1), MoO$_3$ and NaCl (mass ratio 6:1) as the respective precursors.

*Direct transfer of Nb-doped WS$_2$ onto a TEM grid.* Schematics of whole processes are shown in Figure S6. Isopropyl alcohol (IPA) was dropped on the sample which was covered by a TEM grid and was then baked at 70-80 °C for 10 min. HF: H$_2$O (1:6, volume ratio) was used to etch away the SiO$_2$/Si substrate. After etching, the WS$_2$/TEM grid was washed in deionized water and dried for TEM characterization.

*Characterization of the samples*. An optical microscope (Carl Zeiss Microscopy, Germany) and an atomic force microscopy (Cypher ES, Asylum Research, USA) were used to observe the morphology and thickness of the TMDCs. Raman and PL spectroscopy were performed under 532 nm laser excitation (Horiba LabRaman Evolution, Japan). The step size of Raman and PL maps was ~0.8 µm. All the peaks were calibrated using the Si peak at 520.7 cm$^{-1}$. Chemical analyses of the samples were performed by XPS (Thermo Scientific K-Alpha XPS, using Al (Kα) radiation as a probe, USA).

Microstructure characterization of the samples was performed using a STEM (FEI Titan Themis G2 double aberration corrected TEM with a super energy-dispersive X-ray spectroscopy detector, USA) operated at an acceleration voltage of 60 kV.

*FET devices fabrication and measurements.* The Nb-doped WS$_2$ was coated with a PMMA thin film by spin-coating at 3000 r.p.m. for 1 min followed by baking at 125 °C for 30 min. The baked sample was soaked in KOH (1 M) to detach the PMMA/Nb-doped WS$_2$ film from the SiO$_2$/Si substrate, and the detached film was then attached to a fresh 300-nm-thick SiO$_2$/Si substrate and baked at 80 °C for 30 min to promote the adhesion of the Nb-doped WS$_2$ to the substrate. The PMMA film was removed by soaking in acetone overnight. Finally, source and drain electrodes (5/50 nm Cr/Au) were fabricated on the Nb-doped WS$_2$ using a direct laser writing system (miDALIX, DaLI, Slovenia) followed by e-beam evaporation (ebeam) and lift-off processes. The FETs were measured using a semiconductor analyzing system and probe station with vacuum at a base pressure of $10^{-5}$ mbar at room temperature (Keithley 4200A-SCS, USA, LakeShore, USA). According to the electrical measurements, we calculated **carrier concentration (*n*)** and **threshold voltage ($V_\text{th}$)** using the following formulae[20]

$$n = \frac{I_d L}{qWV_d\mu} \quad (1)$$

where $q$ is the unit electron charge, $I_\text{d}$ is the channel current, $V_\text{d}$ is the drain voltage, $\mu$ is the carrier mobility, $L$ and $W$ are the length and width of the channel, respectively.

$$\mu = \frac{L}{WV_\text{d}C_\text{ox}}\frac{dI_\text{d}}{dV_\text{g}} \quad (2)$$

where $C_\text{ox}$ is the gate oxide capacitance evaluated as 11.5 nF·cm$^{-2}$, $I_\text{d}$ is the drain-source current evaluated at the gate voltage ($V_\text{g}$).

*HER-Microreactor fabrication and measurements.* The microreactor was fabricated using two rounds of standard electron beam lithography on the Nb-doped or pure $WS_2$, and Cr/Au source/drain contacts (5/50 nm Cr/Au) were deposited by ebeam. The HER measurements were performed using a three-electrode electrochemical system in a 0.5 M $H_2SO_4$ (aq) electrolyte. A Ag/AgCl electrode was used as the reference microelectrode and a Pt wire (diameter of 1 mm) served as the counter electrode. A gold electrode connected to the Nb-doped $WS_2$ or pure $WS_2$ was used as the working electrode. An electrochemical workstation (Biological VNP300, France) was used for the LSV tests with a scan rate of 0.5 mV s$^{-1}$ to avoid nonfaradic currents. Before testing, the electrolyte was bubbled with Ar for 30-45 min to remove any dissolved oxygen, and the $H_2SO_4$ electrolyte (50 μL) was then dropped onto the device for further measurements. All results were without iR compensation.

*DFT Calculations.* We performed first-principles DFT calculations using the Vienna ab-initio simulation package (VASP),[35-36] where the ion-electron interactions and the exchange correlations were treated by the projected-augmented wave method[37-38] and the Perdew-Burke-Ernzerhof functional[39] for the generalized gradient approximation. A vacuum layer greater than 10 Å thick was used to keep interactions from the neighboring van der Waals layers negligible, while a superlattice larger than 5×5-unit cells was adopted to minimize the influence of neighboring dopants. The cut-off energy was set as 400 eV, while the criteria for energy and force convergence were 10$^{-6}$ eV and 0.01 eV/Å, respectively. A DFT-D3 method with Becke-Jonson damping[40] was used to treat the van der Waals interaction. The zero-point energy and entropy were also considered in calculating the Gibbs free energy difference.

## ASSOCIATED CONTENT

**Supporting Information**

**AUTHOR INFORMATION**

**Corresponding Authors**

*E-mail: bilu.liu@sz.tsinghua.edu.cn

*E-mail: hmcheng@sz.tsinghua.edu.cn

**ORCID**

Bilu Liu: 0000-0002-7274-5752

Hui-Ming Cheng: 0000-0002-5387-4241

**Notes**

The authors declare no competing financial interest.

**ACKNOWLEDGMENTS**

We acknowledge the supports by the National Natural Science Foundation of China (Nos. 51722206, 51920105002, 51991340, and 51991343), the National Key R&D Program (2018YFA0307200), Guangdong Innovative and Entrepreneurial Research Team Program (No. 2017ZT07C341), the Bureau of Industry and Information Technology of Shenzhen for the "2017 Graphene Manufacturing Innovation Center Project" (No. 201901171523). This work is also assisted by SUSTech Core Research Facilities, especially technical support from Pico-Centre that receives support from Presidential fund and Development and Reform Commission of Shenzhen Municipality.

**REFERENCES**

[1] R. Kappera, D. Voiry, S. E. Yalcin, B. Branch, G. Gupta, A. D. Mohite, M. Chhowalla, *Nat. Mater.* **2014**, 13, 1128.

[2] L. Fu, F. Wang, B. Wu, N. Wu, W. Huang, H. Wang, C. Jin, L. Zhuang, J. He, L. Fu, Y. Liu, *Adv. Mater.* **2017**, 29.

[3] Y. He, Q. He, L. Wang, C. Zhu, P. Golani, A. D. Handoko, X. Yu, C. Gao, M. Ding, X. Wang, F. Liu, Q. Zeng, P. Yu, S. Guo, B. I. Yakobson, L. Wang, Z. W. Seh, Z. Zhang, M. Wu, Q. J. Wang, H. Zhang, Z. Liu, *Nat. Mater.* **2019**, 18, 1098.

[4] L. Tang, T. Li, Y. Luo, S. Feng, Z. Cai, H. Zhang, B. Liu, H.-M. Cheng, *ACS Nano* **2020**, 14, 4646.

[5] S. Yuan, C. Shen, B. Deng, X. Chen, Q. Guo, Y. Ma, A. Abbas, B. Liu, R. Haiges, C. Ott, T. Nilges, K. Watanabe, T. Taniguchi, O. Sinai, D. Naveh, C. Zhou, F. Xia, *Nano Lett.* **2018**, 18, 3172.

[6] S. Chen, R. Xu, J. Liu, X. Zou, L. Qiu, F. Kang, B. Liu, H.-M. Cheng, *Adv. Mater.* **2019**, 31, 1804810.

[7] A. Colli, A. Fasoli, C. Ronning, S. Pisana, S. Piscanec, A. C. Ferrari, *Nano Lett.* **2008**, 8, 2188.

[8] P. Luo, F. Zhuge, Q. Zhang, Y. Chen, L. Lv, Y. Huang, H. Li, T. Zhai, *Nanoscale Horiz.* **2019**, 4, 26.

[9] K. Zhang, B. M. Bersch, J. Joshi, R. Addou, C. R. Cormier, C. Zhang, K. Xu, N. C. Briggs, K. Wang, S. Subramanian, K. Cho, S. Fullerton-Shirey, R. M. Wallace, P. M. Vora, J. A. Robinson, *Adv. Funct. Mater.* **2018**, 28.

[10]   R. Xu, X. Zou, B. Liu, H.-M. Cheng, *Mater. Today* **2018**, 21, 391.

[11]   H. Li, X. Duan, X. Wu, X. Zhuang, H. Zhou, Q. Zhang, X. Zhu, W. Hu, P. Ren, P. Guo, L. Ma, X. Fan, X. Wang, J. Xu, A. Pan, X. Duan, *J. Am. Chem. Soc.* **2014**, 136, 3756.

[12]   V. Kochat, A. Apte, J. A. Hachtel, H. Kumazoe, A. Krishnamoorthy, S. Susarla, J. C. Idrobo, F. Shimojo, P. Vashishta, R. Kalia, A. Nakano, C. S. Tiwary, P. M. Ajayan, *Adv. Mater.* **2017**, 29.

[13]   Y. Shi, Y. Zhou, D.-R. Yang, W.-X. Xu, C. Wang, F.-B. Wang, J.-J. Xu, X.-H. Xia, H.-Y. Chen, *J. Am. Chem. Soc.* **2017**, 139, 15479.

[14]   Q. Xiong, Y. Wang, P.-F. Liu, L.-R. Zheng, G. Wang, H.-G. Yang, P.-K. Wong, H. Zhang, H. Zhao, *Adv. Mater.* **2018**, 30, 1801450.

[15]   J. D. Lin, C. Han, F. Wang, R. Wang, D. Xiang, S. Qin, X.-A. Zhang, L. Wang, H. Zhang, A. T. S. Wee, W. Chen, *ACS Nano* **2014**, 8, 5323.

[16]   L. Yang, K. Majumdar, H. Liu, Y. Du, H. Wu, M. Hatzistergos, P. Y. Hung, R. Tieckelmann, W. Tsai, C. Hobbs, P. D. Ye, *Nano Lett.* **2014**, 14, 6275.

[17]   J. Ren, C. Teng, Z. Cai, H. Pan, J. Liu, Y. Zhao, B. Liu, *Sci. China Mater.* **2019**, 62, 1837.

[18]   M. Tosun, L. Chan, M. Amani, T. Roy, G. H. Ahn, P. Taheri, C. Carraro, J. W. Ager, R. Maboudian, A. Javey, *ACS Nano* **2016**, 10, 6853.

[19]   A. Azcatl, X. Qin, A. Prakash, C. Zhang, L. Cheng, Q. Wang, N. Lu, M. J. Kim, J. Kim, K. Cho, R. Addou, C. L. Hinkle, J. Appenzeller, R. M. Wallace, *Nano Lett.* **2016**, 16, 5437.

[20]   B. Tang, Z. G. Yu, L. Huang, J. Chai, S. L. Wong, J. Deng, W. Yang, H. Gong, S. Wang, K. W. Ang, Y. W. Zhang, D. Chi, *ACS Nano* **2018**, 12, 2506.

[21]   J. Suh, T.-E. Park, D.-Y. Lin, D. Fu, J. Park, H. J. Jung, Y. Chen, C. Ko, C. Jang, Y. Sun, R. Sinclair, J. Chang, S. Tongay, J. Wu, *Nano Lett.* **2014**, 14, 6976.

[22]  J. Suh, T. L. Tan, W. Zhao, J. Park, D.-Y. Lin, T.-E. Park, J. Kim, C. Jin, N. Saigal, S. Ghosh, Z. M. Wong, Y. Chen, F. Wang, W. Walukiewicz, G. Eda, J. Wu, *Nat. Commun.* **2018**, 9.

[23]  H. Duan, P. Guo, C. Wang, H. Tan, W. Hu, W. Yan, C. Ma, L. Cai, L. Song, W. Zhang, Z. Sun, L. Wang, W. Zhao, Y. Yin, X. Li, S. Wei, *Nat. Commun.* **2019**, 10, 1584.

[24]  F. Zhang, Y. Lu, D. S. Schulman, T. Zhang, K. Fujisawa, Z. Lin, Y. Lei, A. L. Elias, S. Das, S. B. Sinnott, M. Terrones, *Sci. Adv.* **2019**, 5, eaav5003.

[25]  Q. Li, X. Zhao, L. Deng, Z. Shi, S. Liu, Q. Wei, L. Zhang, Y. Cheng, L. Zhang, H. Lu, W. Gao, W. Huang, C.-W. Qiu, G. Xiang, S. J. Pennycook, Q. Xiong, K. P. Loh, B. Peng, *ACS Nano* **2020**, 14, 4636.

[26]  T. Zhang, K. Fujisawa, F. Zhang, M. Liu, M. C. Lucking, R. N. Gontijo, Y. Lei, H. Liu, K. Crust, T. Granzier-Nakajima, H. Terrones, A. L. Elías, M. Terrones, *ACS Nano* **2020**, 14, 4326.

[27]  F. Cui, Q. Feng, J. Hong, R. Wang, Y. Bai, X. Li, D. Liu, Y. Zhou, X. Liang, X. He, Z. Zhang, S. Liu, Z. Lei, Z. Liu, T. Zhai, H. Xu, *Adv. Mater.* **2017**, 29.

[28]  Q. Deng, X. Li, H. Si, J. Hong, S. Wang, Q. Feng, C. X. Hu, S. Wang, H. L. Zhang, K. Suenaga, H. Xu, *Adv. Funct. Mater.* **2020**, DOI: 10.1002/adfm.202003264.

[29]  V. P. Pham, G. Y. Yeom, *Adv. Mater.* **2016**, 28, 9024.

[30]  Z. Qin, L. Loh, J. Wang, X. Xu, Q. Zhang, B. Haas, C. Alvarez, H. Okuno, J. Z. Yong, T. Schultz, N. Koch, J. Dan, S. J. Pennycook, D. Zeng, M. Bosman, G. Eda, *ACS Nano* **2019**, 13, 10768.

[31]  Y. Jin, Z. Zeng, Z. Xu, Y.-C. Lin, K. Bi, G. Shao, T. S. Hu, S. Wang, S. Li, K. Suenaga, H. Duan, Y. Feng, S. Liu, *Chem. Mater.* **2019**, 31, 3534.

[32]  J. Suh, T. E. Park, D. Y. Lin, D. Fu, J. Park, H. J. Jung, Y. Chen, C. Ko, C. Jang, Y. Sun, R.

Sinclair, J. Chang, S. Tongay, J. Wu, *Nano Lett.* **2014**, 14, 6976.

[33]  Y. Luo, S. Zhang, H. Pan, S. Xiao, Z. Guo, L. Tang, U. Khan, B. F. Ding, M. Li, Z. Cai, Y. Zhao, W. Lv, Q. Feng, X. Zou, J. Lin, H. M. Cheng, B. Liu, *ACS Nano* **2020**, 14, 767.

[34]  J. Greeley, T. F. Jaramillo, J. Bonde, I. Chorkendorff, J. K. Nørskov, *Nat. Mater.* **2006**, 5, 909.

[35]  G. Kresse, J. Furthmüller, *Comput. Mater. Sci.* **1996**, 6, 15.

[36]  G. Kresse, J. Furthmuller, *Phys. Rev. B Condens. Matter.* **1996**, 54, 11169.

[37]  P. E. Blochl, *Phys. Rev. B* **1994**, 50, 17953.

[38]  G. Kresse, D. Joubert, *Phys. Rev. B* **1999**, 59, 1758.

[39]  J. P. Perdew, K. Burke, M. Ernzerhof, *Phys. Rev. Lett.* **1996**, 77, 3865.

[40]  S. Grimme, S. Ehrlich, L. Goerigk, *J. Comput. Chem.* **2011**, 32, 1456.